\documentclass[aps,prl,twocolumn,byrevtex,floatfix,superscriptaddress,longbibliography]{revtex4-1}

\usepackage{amssymb,amsmath,graphicx,afterpage}
\usepackage{hyperref}
\usepackage{color,bm}
\pdfoutput=1

\begin{document}
\title{\boldmath Cooperative cluster-Jahn-Teller effect as a possible route to antiferroelectricity \unboldmath}
\author{K. Geirhos}
\affiliation{Experimental Physics V, Center for Electronic
Correlations and Magnetism, University of Augsburg, 86135 Augsburg,
Germany}
\author{J. Langmann}
\affiliation{CPM, Institute of Physics, University of Augsburg, 86135 Augsburg, Germany}
\author{L. Prodan}
\affiliation{Experimental Physics V, Center for Electronic
Correlations and Magnetism, University of Augsburg, 86135 Augsburg,
Germany} \affiliation{Institute of Applied Physics, MD 2028, Chisinau,  Republic of Moldova}
\author{A. A. Tsirlin}
\affiliation{Experimental Physics VI, Center for Electronic
Correlations and Magnetism, University of Augsburg, 86135 Augsburg,
Germany}
\author{A. Missiul}
\affiliation{'CELLS-ALBA Synchrotron, Cerdanyola del Valles, E-08290 Barcelona, Spain}
\author{G. Eickerling}
\affiliation{CPM, Institute of Physics, University of Augsburg, 86135 Augsburg, Germany}
\author{A. Jesche}
\affiliation{Experimental Physics VI, Center for Electronic
Correlations and Magnetism, University of Augsburg, 86135 Augsburg,
Germany}
\author{V. Tsurkan}
\affiliation{Experimental Physics V, Center for Electronic
Correlations and Magnetism, University of Augsburg, 86135 Augsburg,
Germany} \affiliation{Institute of Applied Physics, MD 2028, Chisinau,  Republic of Moldova}
\author{P. Lunkenheimer}
\affiliation{Experimental Physics V, Center for Electronic
Correlations and Magnetism, University of Augsburg, 86135 Augsburg,
Germany}
\author{W. Scherer}
\affiliation{CPM, Institute of Physics, University of Augsburg, 86135 Augsburg, Germany}
\author{I. K\'ezsm\'arki}
\affiliation{Experimental Physics V, Center for Electronic
Correlations and Magnetism, University of Augsburg, 86135 Augsburg,
Germany}
\begin{abstract}

We report the observation of an antipolar phase in cubic GaNb$_4$S$_8$ driven by an unconventional microscopic mechanism, the cooperative Jahn-Teller effect of Nb$_4$S$_4$ molecular clusters. The assignment of the antipolar nature is based on sudden changes in the crystal structure and a strong drop of the dielectric constant at $T_\mathrm{JT}=31$\,K, also indicating the first-order nature of the transition. In addition, we found that local symmetry lowering precedes long-range orbital ordering, implying the presence of a dynamic Jahn-Teller effect in the cubic phase above $T_\mathrm{JT}$. Based on the variety of structural polymorphs reported in lacunar spinels, also including ferroelectric phases, we argue that GaNb$_4$S$_8$ may be transformable to a ferroelectric state, which would further classify the observed antipolar phase as antiferrolectric.
\end{abstract}

\maketitle

Macroscopic stray fields inherent to ferromagnetic and ferroelectric orders allow coupling to uniform external fields and make these states easy to identify and control. Therefore, ferromagnetic and ferroelectric materials are extensively used in various applications, most prominently in information technology. In contrast, in antiferromagnetic and antiferroelectric (AFE) materials no macroscopic stray fields develop due to the staggered order of  dipoles. Until the development of neutron diffraction, this hindered the direct observation of antiferroic ordering \cite{Shull1949}. However, when it comes to memory applications, the absence of stray fields can be an advantage, leading to the robustness of antiferroic states against unwanted switchings by disturbing macroscopic fields. The great potential of antiferromagnets in information technology has triggered an enormous progress in antiferromagnetic spintronics, a recently emerging field of magnetism \cite{Wadley2016, Jungwirth2016, Jungwirth2018, Lebrun2018}.

Though AFE compounds are also of fundamental interest and can possess similar advantages as their magnetic counterparts, this type of order is much less explored due to conceptual difficulties in defining the basic criteria of antiferroelectricity and identifying its unique experimental signatures. Beside studies on model-type perovskite antiferroelectrics \cite{Hao2014, Liu2011, Hlinka2014, Tagantsev2013, Toledano2019} and antiferroelectricity in liquid crystals \cite{Takezoe2010}, there are only few reports on AFE order in other material classes \cite{Smith2007, Unoki1977, Kishimoto2010, Albers1982, Horiuchi2018, Wu2019, MilesiBrault2020}.

\begin{figure}[t]
\includegraphics[width=\linewidth] {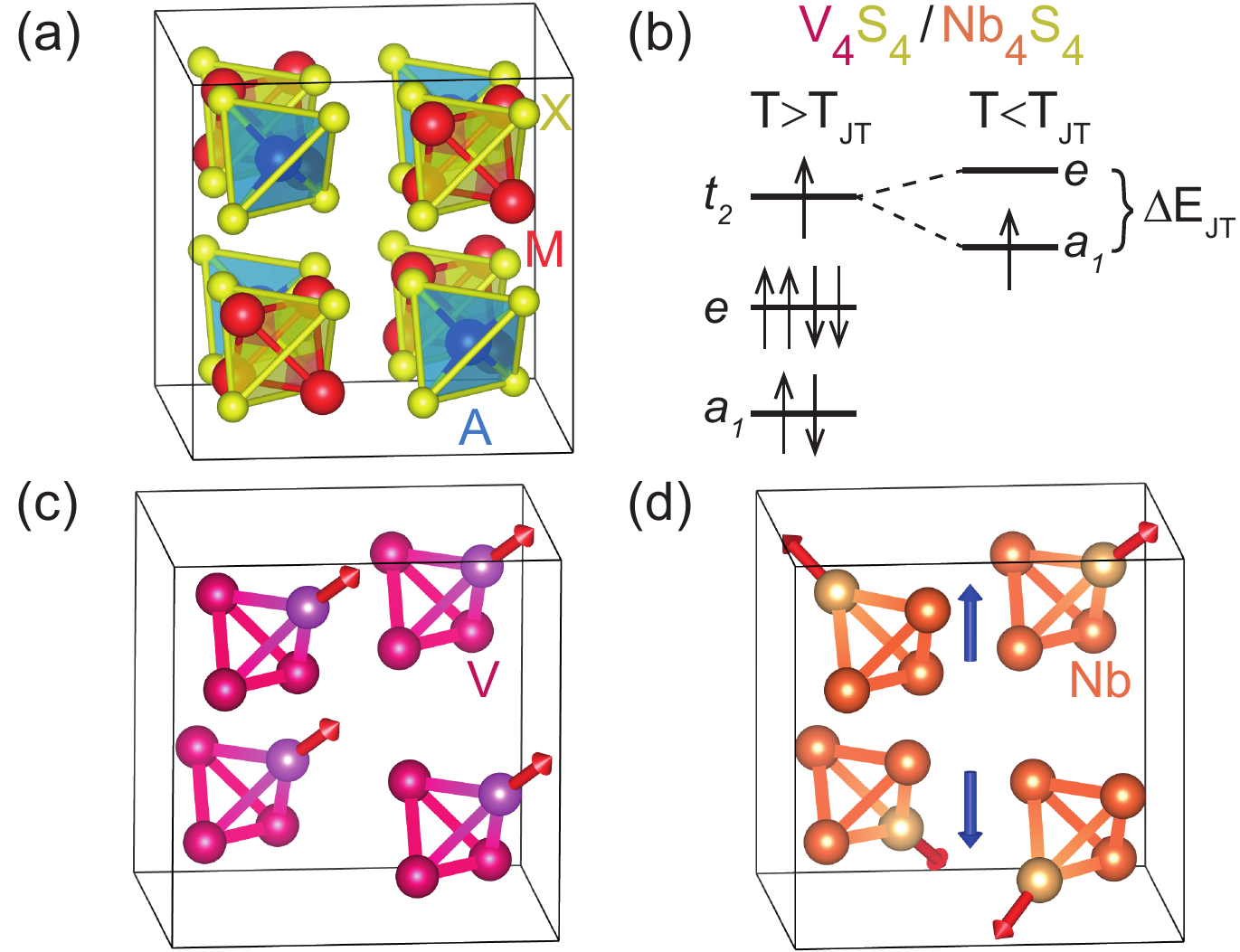}
\caption{\label{Structure} (a) Structural model of the cubic lacunar spinels $AM_4X_8$. (b) Molecular orbital scheme and electronic configuration of $M_4X_4$ clusters for $M=$~Nb, V and $X=$~S in the cubic and the low-temperature distorted phase without considering spin-orbit effects. (c) and (d) Main motif of the Jahn-Teller distortion of the $M_4$ tetrahedra for GaV$_4$S$_8$ and GaNb$_4$S$_8$, respectively. Red arrows indicate the main distortion component for individual $M_4$ units, while blue arrows indicate the sum of the distortions for the upper and lower Nb$_4$ units. }
\end{figure}

Current approaches define antiferroelectricity via its onset: Accordingly, AFE transitions represent a class of symmetry-lowering structural transitions between two non-polar phases, upon which some of the crystallographic sites become polar \cite{Rabe2013, Toledano2016, Toledano2019}. As experimental feature, the transition to the AFE state is associated with a drop of the dielectric constant \cite{Toledano2019}. In terms of symmetry, the onset of antiferroelectricity requires the point group of a symmorphic subgroup of the lower-symmetry (AFE) space group to coincide with the site symmetry of at least one of the sites which become polar upon the transition \cite{Toledano2016}. In addition, an alternative polar distortion of the non-polar high-symmetry phase into a ferroelectric phase is required to render the antipolar phase also antiferroelectric \cite{Rabe2013}. If this ferroelectric state and the AFE state are separated by a sufficiently low energy barrier, a transition between them can be driven by laboratory electric fields.

Here, we investigate the emergence of antiferroelectricity in the $AM_4X_8$ lacunar spinel family, a class of cluster Mott-insulators \cite{Streltsov2017, Reschke2020, Kim2020}. The lacunar spinel structure can be derived from the normal spinel structure $AM_2X_4$ by removing every second $A$-site ion, resulting in a non-centrosymmetric cubic structure with the space group $F\bar{4}3m$. These materials are often considered as molecular crystals with weakly linked $M_4X_4$ and $AX_4$ clusters, as seen in Fig.~\ref{Structure}(a) \cite{Barz1973, Yaich1984, Johrendt1998, Pocha2005}. The electronic configuration of the $M_4X_4$ cubane clusters, within a molecular orbital scheme, leads to an unpaired electron in GaV$_4X_8$ and GaNb$_4X_8$ and an unpaired hole in  GaMo$_4X_8$, occupying a triply degenerate molecular orbital, with $X$ being S or Se (see Fig.~\ref{Structure}(b)) \cite{Pocha2000}. This makes the $M_4X_4$ molecular cluster magnetic and Jahn-Teller active.

The orbital degeneracy of the cubic phase, accordingly, is lifted by a cooperative Jahn-Teller distortion of the $M_4X_4$ units. In case of GaV$_4$S$_8$, GaV$_4$Se$_8$, and GaMo$_4$S$_8$ the ferrodistortive nature of the transition leads to a polar state, making these materials rare examples of orbital-order driven ferroelectrics \cite{Ruff2015, Ruff2017, Neuber2018, Geirhos2018, Xu2015, Barone2015}. For GaV$_4$S$_8$ it was even found that the ferroelectric transition is of order-disorder type, meaning that the cooperative Jahn-Teller distortion below $T_\mathrm{JT}$ is preceded by a dynamic Jahn-Teller effect \cite{Wang2015}.  Upon the non-polar to polar transition, taking place between 40-50\,K depending on the material, all $M_4X_4$ clusters get distorted along the same cubic body diagonal, reducing the symmetry to rhombohedral (space group $R3m$) \cite{Pocha2000, Ruff2015, Ruff2017, Neuber2018}, as seen in Fig.~\ref{Structure}(c). The point group of the crystal coincides with that of the rhombohedrally distorted $M_4X_4$ cubanes ($3m$). Recently, electric and magnetic control of ferroelectric domains \cite{Ghara2020, Ruff2017, Geirhos2020, Neuber2018}, magnetoelectric effects \cite{Ghara2020, Janod2015} and skyrmion lattices \cite{Kezsmarki2015, Fujima2017, Bordacs2017, Zhang2019} have been reported in these compounds.

In contrast, in GaNb$_4$S$_8$ the orbital degeneracy of the Nb$_4$S$_4$ units is lifted by a more complex distortion, which leads to a stronger symmetry reduction of the Nb$_4$S$_4$ units to point group $m$ \cite{Jakob2007}. Still, similar to its ferrodistortive sister compounds, the main motif in the Jahn-Teller driven deformation of individual Nb$_4$S$_4$ clusters is the rhombohedral distortion, i.e. an elongation parallel to one of the cubic body diagonals, inducing polar moments on the Nb$_4$S$_4$ units. However, the cooperative distortion of four adjacent Nb$_4$S$_4$ units below $T_\mathrm{JT}$ is along four different cubic body diagonals, as sketched in Fig.~\ref{Structure}(d), thus, preventing the development of a macroscopic polarization and leading to an overall tetragonal symmetry  \cite{Jakob2007}. Interestingly, this tetragonal distortion in GaNb$_4$S$_8$ from space group $F\bar{4}3m$ to space group $P\bar{4}2_1m$  fulfills the group-theoretical criterion for AFE transitions \cite{Toledano2016, Toledano2019}: Ga is the only atom occupying a Wyckoff position with non-polar site symmetry $\bar{4}3m$ in the cubic phase. Upon the transition to the tetragonal phase, a polar site symmetry $m$ is obtained for Ga that coincides with the point group of the symmorphic subgroups $C1m1$ and $P1m1$ of $P\bar{4}2_1m$. A symmetry-mode decomposition analysis of the structural transition also revealed that the distortion is dominated by the Jahn-Teller active $X_5$ mode that lifts the orbital degeneracy of the Nb$_4$ units (see Table~S2 in the supplemental material \cite{Suppl} including Refs.~\cite{Reisinger2007, Langmann2019, Duisenberg1992, Duisenberg2003, Fauth2014, jana2006, Varignon2015, Varignon2019, Stokes, Stokes2006, Jakob2007, Singh2014, Ruff2015, Ruff2017, Neuber2018, Kohara2010, Nhalil2015, Dai1995, Nagai2019, Constable2017, Wei2014, Schranz2019}).

\begin{figure}[t]
\includegraphics[width=1\linewidth] {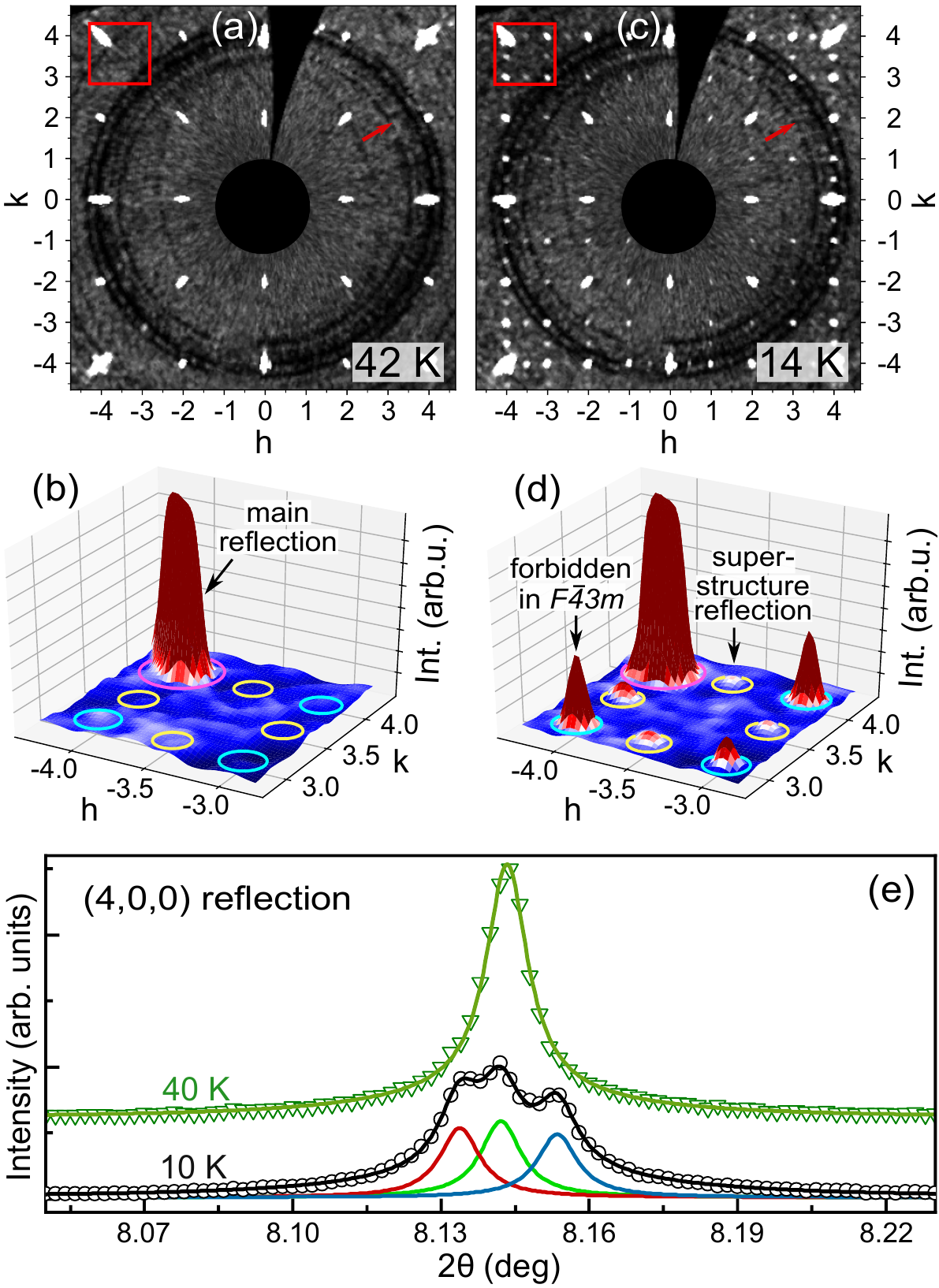}
\caption{\label{XRD} Reconstructions of the $(hk0)$ reciprocal-space plane measured at 42\,K (a) and 14\,K (c). An overview of the different reflection types visible at 42\,K and 14\,K is given by three-dimensional plots (b, d) of the regions marked by red rectangles in (a) and (c). Intense main reflections in (b) and (d) are clipped. Ring-shaped features marked by red arrows are due to parasitic scattering. Panel (e) shows the fine structure  of the (4,\,0,\,0) reflection measured by powder XRD at 40\,K and 10\,K. The solid black and dark green lines represent fits of the data. A decomposition of the threefold-split reflection at 10\,K is indicated by solid red, green and blue lines.}
\end{figure}

To further investigate the nature of the Jahn-Teller transition and the possible emergence of antiferroelectricity in GaNb$_4$S$_8$ we apply a multi-probe approach, including dielectric spectroscopy, single-crystal and high-resolution powder X-ray diffraction (XRD), specific heat, magnetic suscceptibility measurements. Our results evidence that the orbital degeneracy of Nb$_4$S$_4$ clusters drives a first-order transition to a non-magnetic antipolar phase at $T_{\mathrm{JT}}=31$\,K. The powder XRD measurements, in addition, also suggest the presence of a dynamic Jahn-Teller effect in the cubic phase above $T_\mathrm{JT}$.

All experiments, except the powder XRD, were performed on single-crystals grown by the chemical transport reaction method. (For experimental details see the supplemental material \cite{Suppl}.)

Figs.~\ref{XRD}(a) and (c) display reconstructions of the $(hk0)$ reciprocal-space plane from single-crystal XRD experiments at 42\,K and 14\,K, respectively. (For extended and additional reciprocal-space planes see Fig.~S2 \cite{Suppl}.) Below $T_\mathrm{JT}=31$\,K, diffracted intensity was detected at positions of previously forbidden reflections of the type $hkl: h+k, h+l, k+l = 2n+1$ (see Figs.~\ref{XRD}(b) and (d)). This is compatible with the symmetry lowering from the cubic space group $F\bar{4}3m$ ($a=9.9837(2)$\,\AA) to the tetragonal space group $P\bar{4}2_1m$ reported earlier \cite{Jakob2007}. Interestingly, we found additional weak reflections at half-integer positions indicating a doubling of one unit cell axis (Fig.~\ref{XRD}(d)).

To detect potential small changes in the cell metrics we performed synchrotron powder XRD experiments at ALBA. As shown in Fig.~\ref{XRD}(e), below $T_\mathrm{JT}$ the (4,\,0,\,0) reflection splits into three peaks with nearly equal intensities, indicating that the symmetry of the low-temperature phase is not higher than orthorhombic. (The tetragonal $P\bar 42_1m$ symmetry would lead to a splitting of (4,\,0\,,0) into two peaks.) Both single-crystal and powder XRD data are consistent with a description of the low-temperature structure in the non-polar but chiral orthorhombic space group $P2_12_12_1$ ($a = 9.9873(4)$\,\AA , $b = 9.9749(3)$\,\AA , $c = 19.9775(6)$\,\AA; for a comparison of the crystallographic data above and below $T_\mathrm{JT}$ see Table~S1 \cite{Suppl}). This space group choice takes into account the orthorhombicity, the extinction condition  $h00, h=2n$, $0k0, k=2n$ and  $00l, l=2n$  and a unit-cell doubling along the $c$-axis. Very low intensities of the superstructure reflections, unfortunately, do not allow the extraction of the detailed distortion pattern. However, the low intensities also suggest that the distortion should be dominated by the same $X_5$ mode as for the $P\bar 42_1m$ model and, thus, reflect the Jahn-Teller instability. Sill it is clear, that the remaining mirror-plane symmetry of the Nb$_4$ clusters in space group $P\bar{4}2_1m$ is lost in space group $P2_12_12_1$. Correspondingly, the site symmetry of Ga, Nb and S is reduced to polar 1 coinciding with the point group of the symmorphic polar subgroup $P1$ of $P2_12_12_1$. Therefore, the transition from $F\bar{4}3m$ to $P2_12_12_1$ also fulfills the symmetry criterion for AFE transitions \cite{Toledano2016}.

\begin{figure}[t]
\includegraphics[width=\linewidth] {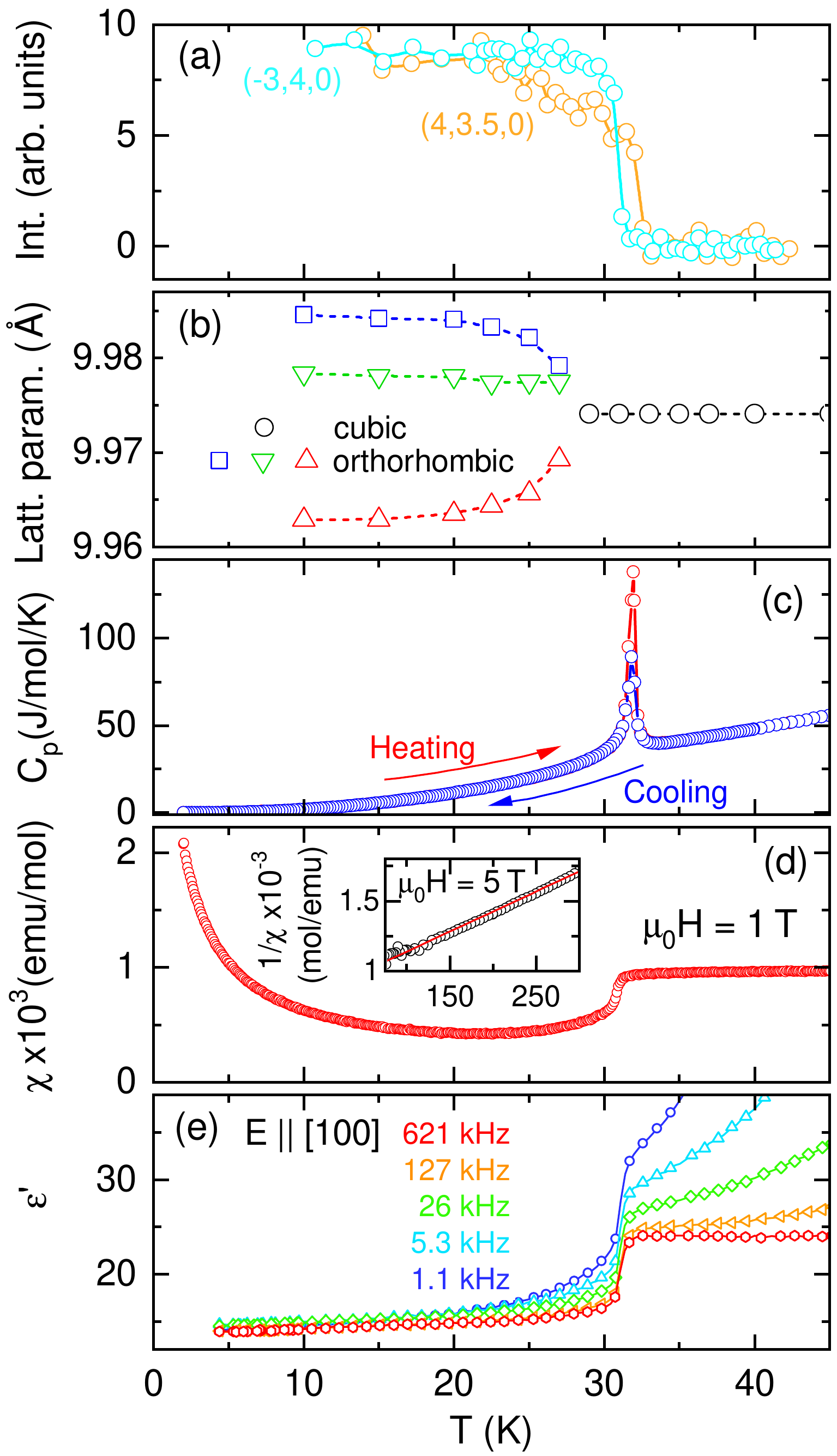}
\caption{\label{Tdep}Temperature dependence of (a) the reflection intensities at positions (-3,\,4,\,0) and (4,\,3.5,\,0) as recorded by single-crystal XRD (rescaled to the same intensity level at low temperatures), (b) the lattice parameters, (c) specific heat, (d) magnetic susceptibility, and (e) dielectric constant at different frequencies. The inset in panel (d) displays the inverse susceptibility at elevated temperatures.}
\end{figure}

To determine the temperature dependence of the structural distortion, we recorded the intensity of the (-3,\,4\,,0) reflection, forbidden in the high-temperature space group, and the intensity of the (4,\,3.5,\,0) reflection, indicating the cell doubling (Fig.~\ref{Tdep}~(a)). With decreasing temperature, a sudden appearance of intensity is observed at $T_\mathrm{JT}$, followed by a weak gradual increase towards lower temperatures. This points to a first-order character of the transition. An abrupt change of the lattice parameters and the cell volume as determined by powder XRD measurements (Fig.~\ref{Tdep}(b) and Fig.~S3 \cite{Suppl}) further supports the first-order nature. We also observe higher peak values of the $\lambda$-shaped anomaly of specific heat on heating than on cooling (Fig.~\ref{Tdep}~(c)), characteristic for first-order transitions, caused by the latent heat associated with the transition. In the lacunar spinels undergoing a ferroelastic-ferroelectric distortion, the transition was also found to be of weakly first order \cite{Singh2014, Ruff2015, Ruff2017, Geirhos2018}.

As a key result of this work, the temperature dependence of the dielectric constant in Fig.~\ref{Tdep}(e) reveals a discontinuous, $\sim$30\% decrease at $T_\mathrm{JT}$. The structural transition, therefore, does not only fulfill the symmetry criterion for an AFE transition, but this instantaneous drop of the dielectric constant also agrees with the theoretical expectations for a first-order AFE transition \cite{Toledano2016, Toledano2019}. Though the existence of and the possible switching to an alternative polar state will be discussed later, we tentatively refer to the antipolar state as an AFE state. The jump is seen at each frequency, though the intrinsic plateau-like behavior of $\varepsilon'$ above $T_\mathrm{JT}$ is masked by a gradual increase observed for low frequencies due to extrinsic Maxwell-Wagner relaxation \cite{Lunkenheimer2002, Lunkenheimer2009}. This behavior of $\varepsilon'$ is distinct from that of the polar sister compounds, where $\varepsilon'$ exhibits a peak  at $T_\mathrm{JT}$, characteristic of ferroelectric ordering \cite{Ruff2015, Fujima2017, Geirhos2018}. GaNb$_4$S$_8$ is, therefore, the first material, where the Jahn-Teller activity of molecular clusters has been identified as a potential driving mechanism of AFE ordering.

\begin{figure}
\includegraphics[width=\linewidth] {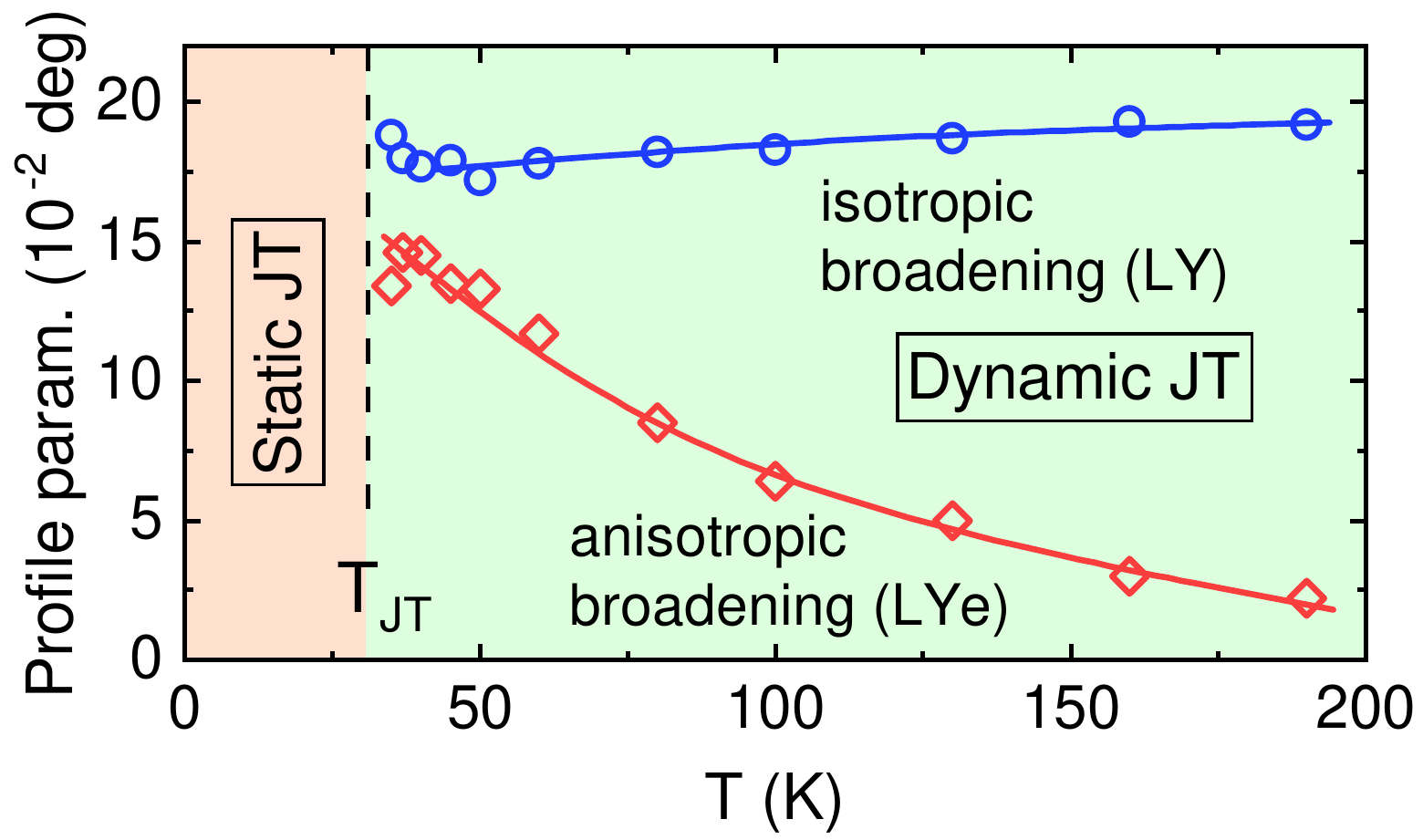}
\caption{\label{Dynamic} Isotropic (LY) and anisotropic (LYe) contributions to the peak width by powder XRD above $T_{\rm JT}$. The lines are guides to the eye.}
\end{figure}

Remarkably, our powder XRD studies also reveal that local symmetry lowering precedes the cooperative Jahn-Teller transition pointing to a dynamic Jahn-Teller effect in the cubic phase. This is evidenced by an anisotropic peak broadening. The broadening can be decomposed into isotropic and anisotropic contributions using the parameters LY and LYe, respectively. As can be seen in Fig.~\ref{Dynamic}, the isotropic part LY does not change significantly upon cooling, whereas the anisotropic part LYe increases as the temperature approaches $T_{\rm JT}$ from higher temperatures. The observation of a dynamic Jahn-Teller effect is also in agreement with the enhanced structural fluctuations observed by nuclear magnetic resonance well above $T_{\rm JT}$~\cite{Waki2010}, which classifies this transition as order-disorder type.

The structural transition has a strong impact not only on the polar but also on the magnetic state of GaNb$_4$S$_8$. The sudden drop of the magnetic susceptibility at $T_\mathrm{JT}$, shown Fig.~\ref{Tdep}(d), implies that the first-order structural transition is accompanied by spin-singlet formation, as concluded from nuclear magnetic resonance studies \cite{Waki2010}. The antiferromagnetic exchange is clearly manifested in the large, negative Curie-Weiss temperature of $\theta=-300$\,K, as determined from the fit of the inverse susceptibility in the inset of Fig.~\ref{Tdep}(d) \cite{Pocha2005, Jakob2007}. A similar interlocking between the structural transition and the onset of a non-magnetic state has been recently observed in GaNb$_4$Se$_8$ and GaTa$_4$Se$_8$ \cite{Ishikawa2020}. In these compounds, the spin-orbit coupling, in interplay with the Jahn-Teller effect of the $M_4$Se$_4$ clusters, was identified as the driving force of the magneto-structural transition. While crystal-field effects usually dominate over spin-orbit coupling in semiconductors based on $3d$ transition metals, spin-orbit coupling has an increasing impact on the electronic structure of $4d$ and $5d$ compounds. Thus, spin-orbit coupling has to be taken into account for the proper description of the cluster orbitals in these compounds and the microscopic mechanism behind their magnetostructural transitions. In fact, in GaNb$_4$Se$_8$ and GaTa$_4$Se$_8$ the relevance of spin-orbit coupling is clearly manifested in the strong reduction of their effective paramagnetic moments from the S=1/2 spin-only value. However, for GaNb$_4$S$_8$ we obtained an effective moment of $\mu_{\mathrm{eff}}=1.67\,\mu_{\mathrm{B}}/f.u.$ from susceptibility measurements, which is very close to the spin-only value ($1.73\,\mu_{\mathrm{B}}/f.u.$). This implies the secondary role of spin-orbit physics and the dominance of the Jahn-Teller active Nb$_4$ cluster in driving the structural transition.

In addition, we observed a very weak polarization below $T_\mathrm{JT}$ (see Figs.~S4 \cite{Suppl}). Such a weak polarization superimposed on AFE order was previously observed in several materials and ascribed to different origins, like polar domain walls or tiny canting of the antipolar order \cite{Dai1995, Nagai2019, Constable2017}. Since a detailed analysis is beyond the scope of this paper, a discussion of the weak polarization is provided in the supplemental material \cite{Suppl}.

Unfortunately, we did not succeed with transforming the antipolar state to a polar state by electric fields. In this respect, the narrow-gap ($\sim0.2$\,eV) and the lower resistivity of the material, compared to other oxide antiferroelectrics \cite{Liu2011, Cross1955, Uppuluri2019}, introduce strong limitations on the magnitude of static electric fields applicable. Non-resonant photo-induced switching via the strong electric field of intense THz radiation could be one approach to overcome this limitation. In the following we argue that GaNb$_4$S$_8$ is a potential AFE material, i.e. it likely has a polar phase with energy close to that of the antipolar ground state. While lacunar spinels share a cubic structure at room temperature, they realize a variety of polymorphs below the structural transition. Prime examples are GaV$_4$S$_8$, GaTa$_4$Se$_8$ and GaNb$_4$Se$_8$ with a polar rhombohedral ($R3m$), antipolar tetragonal ($P\bar{4}m2$) and chiral orthorhombic ($P2_12_12_1$) ground state, respectively \cite{Pocha2000, Ishikawa2020}.  GaNb$_4$Se$_8$ even has a chiral cubic phase ($P2_13$) at intermediate temperatures \cite{Ishikawa2020}. Despite this structural diversity, their common aspect is that the distortion of individual $M_4X_4$ units is well approximated by a rhombohedral distortion, as schematically shown in Fig.~\ref{Structure} for GaV$_4$S$_8$ and GaNb$_4$S$_8$. The primary role of the single-cluster distortion in the structural transformation is clear from the hierarchy of energy scales. The Jahn-Teller splitting ($\Delta E_\mathrm{JT}$) arising from the distortion of individual clusters is larger than the charge gap, since the corresponding electric-dipole excitation was not observed in the optical conductivity within the gap \cite{Reschke2020}. This implies the presence of a dynamic Jahn-Teller effect well above $T_\mathrm{JT}$, as was indeed reported for GaV$_4$S$_8$ and is also evidenced for GaNb$_4$S$_8$ by our powder XRD measurements (see Fig.~\ref{Dynamic}) \cite{Wang2015, Waki2010}.

Since the structural transition takes place at much lower temperatures ($\mathrm{k}_\mathrm{B} T_\mathrm{JT}\sim 3\, \mathrm{meV} \ll\Delta E_\mathrm{JT}\sim 300\,\mathrm{meV}$), the cooperative long-range ordering must be governed by weak interactions, such as inter-cluster interactions or strain. As $T_\mathrm{JT}$ varies between 30-50\,K for all these compounds, the energy difference between the polymorphs, including the polar and the antipolar states, should be less than $\mathrm{k}_\mathrm{B} T_\mathrm{JT}$. Thus, we believe that switching between the polymorphs via the proper conjugate fields should in principle be possible. Unfortunately, ab initio calculations may not be able to provide a realistic estimate for the energy difference between the antipolar and polar states of GaNb$_4$S$_8$. Due to the interplay of various factors (narrow gap, interplay between on-cluster correlations and spin-orbit effects, sensitivity of inter-cluster hopping to the orbital character of the single-cluster states), DFT+U schemes have not been able to reproduce the finite gap in the cubic state of the material (see \cite{Reschke2020} and references therein). Thus, to determine the energies of the polymorphs with a few meV precision is a great challenge for theory. These fundamental aspects of the correlated cluster-insulator GaNb$_4$S$_8$, distinguish this compound from DyVO$_4$, where the cooperative Jahn-Teller effect caused by Dy$^{3+}$ ions was identified as the origin of AFE order.

In summary, we have performed structural, dielectric, specific heat and magnetic studies to elucidate the polar and magnetic state of the lacunar spinel GaNb$_4$S$_8$ below its structural transition at $T_{\mathrm{JT}}=31$\,K. Our combined single-crystal and powder XRD measurements revealed a unit-cell doubling along the $c$-axis and a structural distortion compatible with space-group $P2_12_12_1$ below $T_\mathrm{JT}$. We demonstrate that the transition leads to the transformation of the non-polar cubic state to an antipolar state, driven by the Jahn-Teller distortion of Nb$_4$S$_4$ molecular clusters. In addition, we also found evidence for a dynamic Jahn-Teller effect preceding the long range orbital-ordering. In contrast to its ferroelectric sister compounds, GaNb$_4$S$_8$  is the first example indicating that cluster Jahn-Teller effect can also provide a non-canonical mechanism for the emergence of antiferroelectricity.

This work was supported by the Deutsche Forschungsgemeinschaft through the Transregional Collaborative Research Center TRR 80 and via the research project Grant No. KE 2370/3-1. We also acknowledge the support by the project ANCD 20.80009.5007.19 (Moldova). L.P. acknowledges support of Deutscher Akademischer Austauschdienst (DAAD).

\end{document}